\documentstyle[preprint,aps,epsfig]{revtex}
\begin{document}
\draft
\title{Interparticle force in polydisperse electrorheological fluids: \\
       Beyond the dipole approximation}
\author{Y. L. Siu, Jones T. K. Wan and K. W. Yu}
\address{Department of Physics, The Chinese University of Hong Kong, \\
         Shatin, New Territories, Hong Kong, China}
\maketitle

\begin{abstract}
We have developed a multiple image method to compute the interparticle 
force for a polydisperse electrorheological (ER) fluid.
We apply the formalism to a pair of dielectric spheres of different 
dielectric constants and calculate the force as a function of 
the separation.
The results show that the point-dipole (PD) approximation errs 
considerably because many-body and multipolar interactions are ignored.
The PD approximation becomes even worse when the dielectric contrast
between the particles and the host medium is large.
From the results, we show that the dipole-induced-dipole (DID) model 
yields very good agreements with the multiple image results for a wide 
range of dielectric contrasts and polydispersity. 
The DID model accounts for multipolar interaction partially and 
is simple to use in computer simulation of polydisperse ER fluids.
\end{abstract}
\vskip 5mm

\section{Introduction}

Polydisperse electrorheological (ER) fluids have attracted considerable
interest recently because the size distribution and dielectric properties
of the suspending particles can have significant impact on the ER response
\cite{Ota}.
Real ER fluids must be polydisperse in nature:
the suspending particles can have various sizes or different permittivities.
In a recent paper \cite{Yu2000}, we investigated the case when the 
suspending particles are of different sizes. 
In this work, we extend the study to that of different dielectric constants.

The point-dipole approximation \cite{RTao} is routinely adopted in 
computer simulation \cite{Kling89,Kling91} 
because it is simple and easy to use. 
Since many-body and multipolar interactions between particles have been 
neglected, the predicted strength of ER effects is of an order lower than 
the experimental results. 
Hence, substantial effort has been made to sort out more accurate models
\cite{Kling,Davis,Clercx}.
Recently, we have developed a multiple image method and an integral equation 
approach to compute the interparticle force.
In particular, we proposed a dipole-induced-dipole (DID) model for 
efficient computer simulation of polydisperse ER fluids \cite{Yu2000}.

Poladian \cite{Poladian} claimed that the multiple image method can be 
used to calculate the dipole moment of a pair identical dielectric 
spheres in an applied electric field.
In Ref.\cite{Yu2000}, we generalized the multiple image method to a pair
of dielectric spheres of different sizes.
We showed that the generalization yields a reasonable approximation when 
the spheres have a large dielectric constant. The multiple image method was 
widely adopted \cite{dyer,ijmpb,nler-pre}. The approximation is reasonable
because in ER fluids, the dielectric constant of the particles can be much
larger than that of the host fluid.  
However, the results for low contrast are questionable.

In fact there is a more complicated image method for a dielectric sphere 
\cite{Choy}, which gives the exact image dipole moment of a 
dielectric sphere that placed in front of a point dipole. 
We thus modify the multiple image formula. 
The results of the improved formula agree with the numerical solution 
of an integral equation method \cite{etopim1,etopim2} 
even when the dielectric contrast of the spheres is low. 

In this work, we extend the multiple image method to compute the
interparticle forces for a polydisperse mixture of dielectric spheres
of different dielectric constants.
The DID model will be compared with the Klingenberg's empirical force
expressions \cite{Kling}.

\section{Improved Multiple Image Method}

Here we briefly review the method and extend the method slightly to handle 
different dielectric constants.
Consider a pair of dielectric spheres, of radii $a$ and $b$, 
dielectric constants $\epsilon_1$ and $\epsilon_1'$ respectively, 
separated by a distance $r$. 
The spheres are embedded in a host medium of dielectric constant 
$\epsilon_2$. Upon the application of an electric field ${\bf E}_0$,
the induced dipole moment inside the spheres are respectively given by:
\begin{equation}
p_{a0}=\beta \epsilon_2 E_0 a^3,\ \ \ \
  p_{b0}=\beta'\epsilon_2 E_0 b^3,
\end{equation}
where the dipolar factors $\beta,\beta'$ are given by:
\begin{equation}
\beta={\epsilon_1-\epsilon_2 \over \epsilon_1+2\epsilon_2},\ \ \
\beta'={\epsilon_1'-\epsilon_2 \over \epsilon_1'+2\epsilon_2}.
\end{equation}

From the multiple image method \cite{Yu2000}, 
the total dipole moment inside sphere $a$ is:
\begin{eqnarray}
p_{aT} &=& (\sinh \alpha)^3 \sum_{n=1}^\infty \left[
  {p_{a0} b^3 (-\beta)^{n-1}(-\beta')^{n-1}
    \over (b\sinh n\alpha + a\sinh (n-1)\alpha)^3}
    +{p_{b0} a^3 (-\beta)^{n}(-\beta')^{n-1} 
    \over (r \sinh n\alpha)^3 } \right],
\label{trans-a-dielectric} \\
p_{aL} &=& (\sinh \alpha)^3 \sum_{n=1}^\infty \left[
  {p_{a0} b^3 (2\beta)^{n-1}(2\beta')^{n-1}
  \over (b\sinh n\alpha + a\sinh (n-1)\alpha)^3}
    +{p_{b0} a^3 (2\beta)^{n}(2\beta')^{n-1} 
    \over (r \sinh n\alpha)^3 } \right],
\label{long-a-dielectric} 
\end{eqnarray}
where the subscripts $T$ ($L$) denote a transverse (longitudinal) field, 
i.e., the applied field is perpendicular (parallel) to the line joining 
the centers of the spheres. 
Similar expressions for the total dipole moment inside sphere $b$ can be
obtained by interchanging $a$ and $b$, as well as $\beta$ and $\beta'$. 
The parameter $\alpha$ satisfies:
$$
\cosh\alpha={r^2 - a^2 - b^2 \over 2 ab}.
$$
In Ref.\cite{Yu2000}, we checked the validity of these expressions by 
comparing with the integral equation method. 
We showed that these expression are valid at high contrast. 
Our improved expressions will be shown to be good at low contrast as well 
(see below).

The force between the spheres is given by \cite{Jackson}:
\begin{equation}
F_T = {E_0 \over 2}{\partial\over\partial r}(p_{aT}+p_{bT}),\ \ \ \
F_L = {E_0 \over 2}{\partial\over\partial r}(p_{aL}+p_{bL}).
\end{equation}

For monodisperse ER fluids, Klingenberg defined an empirical force 
expression \cite{Kling}:
\begin{eqnarray}
{{\bf F}\over F_{PD}} = (2F_\parallel \cos^2 \theta - F_\perp \sin^2 \theta)
  \hat{\bf r} + F_\Gamma \sin 2\theta \hat{\theta},
\end{eqnarray}
being normalized to the point-dipole force $F_{PD}=-3 p_{0}^2 /r^4$,
where $F_\parallel, F_\perp$ and $F_\Gamma$ (all tending to unity at large
separation) are three force functions being determined from numerical
solution of Laplace's equation.
The Klingenberg's force functions can be shown to relate to
our multiple image moments as follow (here $a=b$, $\beta=\beta'$ and 
$p_a=p_b$): 
\begin{eqnarray}
F_\parallel={1 \over 2}{\partial \tilde{p}_L \over \partial r}, \ \ \
F_\perp=-{\partial \tilde{p}_T \over \partial r}, \ \ \
F_\Gamma={1\over r}(\tilde{p}_T - \tilde{p}_L),
\label{Klingen}
\end{eqnarray}
where $\tilde{p}_L=p_L/F_{PD}E_0$ and $\tilde{p}_T=p_T/F_{PD}E_0$ are
the reduced multiple image moments. We computed the numerical values of
these force functions separately by the approximant of Table I of the second
reference of Ref.\cite{Kling} and by Eq.(\ref{Klingen}). 

In Fig.1, we plot the multiple image results and the Klingenberg's empirical 
expressions. We show results for the perfectly conducting limit ($\beta=1$) 
only. For convenience, we define the reduced separation $\sigma=r/(a+b)$.
For reduced separation $\sigma>1.1$, simple analytic expressions were 
adopted by Klingenberg. As evident from Fig.1, the agreement with the 
multiple image results is impressive at large reduced separation 
$\sigma>1.5$, for all three empirical force functions. 
However, significant deviations occur for $\sigma<1.5$, especially for 
$F_\parallel$.
For $\sigma\le 1.1$, alternative empirical expressions were adopted by 
Klingenberg. For $F_\perp$, the agreement is impressive, 
although there are deviations for the other two functions.
From the comparison, we would say that reasonable agreements have been 
obtained. Thus, we are confident that the multiple image expressions give
reliable results. 

\section{Dipole-induced-dipole model}

The analytic multiple image results can be used to compare among the 
various models according to how many terms are retained in the
multiple image expressions:
(a) point-dipole (PD) model: $n=1$ term only,
(b) dipole-induced-dipole (DID) model: $n=1$ to $n=2$ terms only, and
(c) multipole-induced-dipole (MID) model: $n=1$ to $n=\infty$ terms.

The multiple image expressions [Eqs.(3)--(6)] allows us to calculate 
the correction factor defined as the ratio between the DID and PD forces: 
\begin{eqnarray}
{F_{DID}^{(T)} \over F_{PD}^{(T)}}
&=& 1 - {\beta a^3 r^5\over (r^2-b^2)^4} - {\beta' b^3 r^5\over (r^2-a^2)^4}
  + {\beta \beta' a^3 b^3 (3r^2-a^2-b^2)\over (r^2-a^2-b^2)^4}, \\
{F_{DID}^{(L)} \over F_{PD}^{(L)}}
&=& 1 + {2\beta a^3 r^5\over (r^2-b^2)^4} + {2\beta' b^3 r^5\over (r^2-a^2)^4}
  + {4\beta \beta' a^3 b^3 (3r^2-a^2-b^2)\over (r^2-a^2-b^2)^4},
\end{eqnarray}
where $F_{PD}^{(T)}=3 p_{a0} p_{b0}/r^4$ and 
$F_{PD}^{(L)}=-6 p_{a0} p_{b0}/r^4$ are the point-dipole forces for the 
transverse and longitudinal cases respectively.
These correction factors can be readily calculated in computer simulation
of polydisperse ER fluids.
The results show that the DID force deviates significantly from the PD force 
at high contrast when $\beta$ and $\beta'$ approach unity. 
The dipole induced interaction will generally decrease (increase) the 
magnitude of the transverse (longitudinal) interparticle force with 
respect to the point-dipole limit.

In a previous work \cite{Yu2000}, we examine the case of different size
but equal dielectric constant ($\beta=\beta'$) only.
Here we focus on the case $a=b$ and study the effect of different 
dielectric constants. 
In Fig.2, we plot the interparticle force in the transverse field case 
against the reduced separation $\sigma$ between the spheres for $\beta=1/3$
and various $\beta'/\beta$ ratios. At low contrast, the DID model almost 
coincides with the MID results.
In contrast, the PD model exhibits significant deviations.
Similar conclusion can be drawn from the longitudinal field case as in Fig.3.
It is evident that the DID model generally gives better results than PD for
all polydispersity.

At higher contrast, the DID model still agrees with the MID model,
except at close encounter. In Figs.4 and 5, we plot the force in the 
transverse and longitudinal field cases against the reduced separation 
$\sigma$. It is evident that the DID model generally gives better results 
than PD for all polydispersity.
For the longitudinal field case, the DID model agrees with the MID model 
for $\sigma > 1.2$, except at close encounter where the MID force 
diverges as $\sigma \to 1$.

\section*{Conclusion}

In summary, we have used the multiple image to compute the interparticle 
force for a polydisperse electrorheological fluid. 
We apply the formalism to a pair of spheres of different dielectric 
constants and calculate the force as a function of the separation.
The results show that the point-dipole approximation is oversimplified. 
It errs considerably because many-body and multipolar interactions are 
ignored.
The dipole-induced-dipole model accounts for multipolar interactions 
partially and yields overall satisfactory results in computer simulation 
of polydisperse ER fluids while it is easy to use.

\section*{Acknowledgements}

This work was supported by the Research Grants Council of the Hong Kong 
SAR Government under grant CUHK4284/00P.

\vskip 5mm\noindent
\begin{center}
{\large Figure Captions}
\end{center}
Fig.1: Comparison of the multiple image results with Klingenberg's empirical
 force expression. \\
Fig.2: Interparticle force for transverse field, $\beta$=1/3 while 
 $\beta'/\beta$ ranges from 1.0 to 1.2. \\
Fig.3: Same as Fig.2. Force for longitudinal field. \\
Fig.4: Force for transverse field, $\beta$=9/11 while 
 $\beta'/\beta$ ranges from 1.0 to 1.2. \\
Fig.5: Same as Fig.4. Force for longitudinal field.

\newpage
\centerline{Fig.1: Comparison with Klingenberg}
\centerline{\epsfig{file=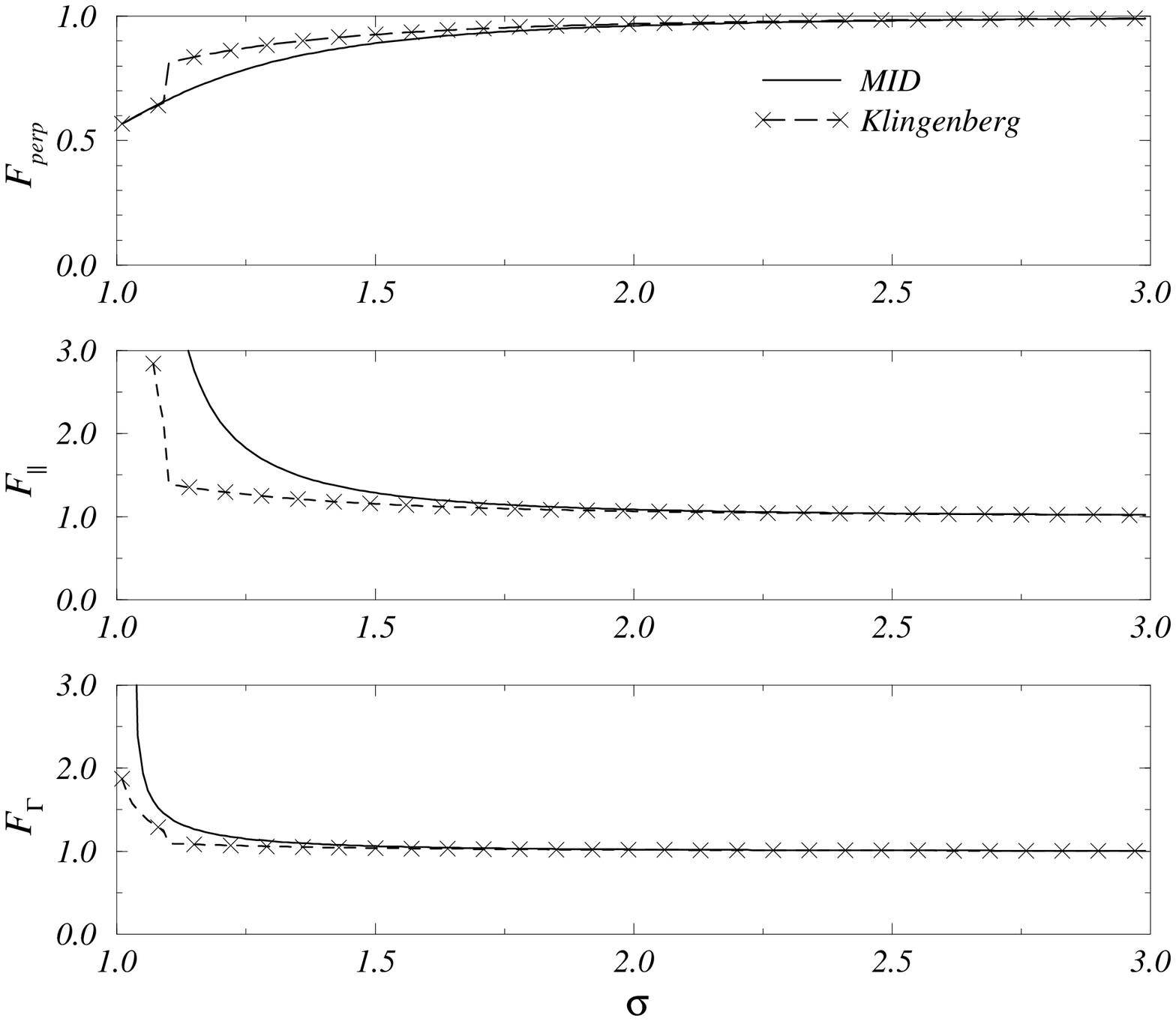,width=\linewidth}}

\newpage
\centerline{Fig.2: Force for Transverse Field, $\beta$=1/3}
\centerline{\epsfig{file=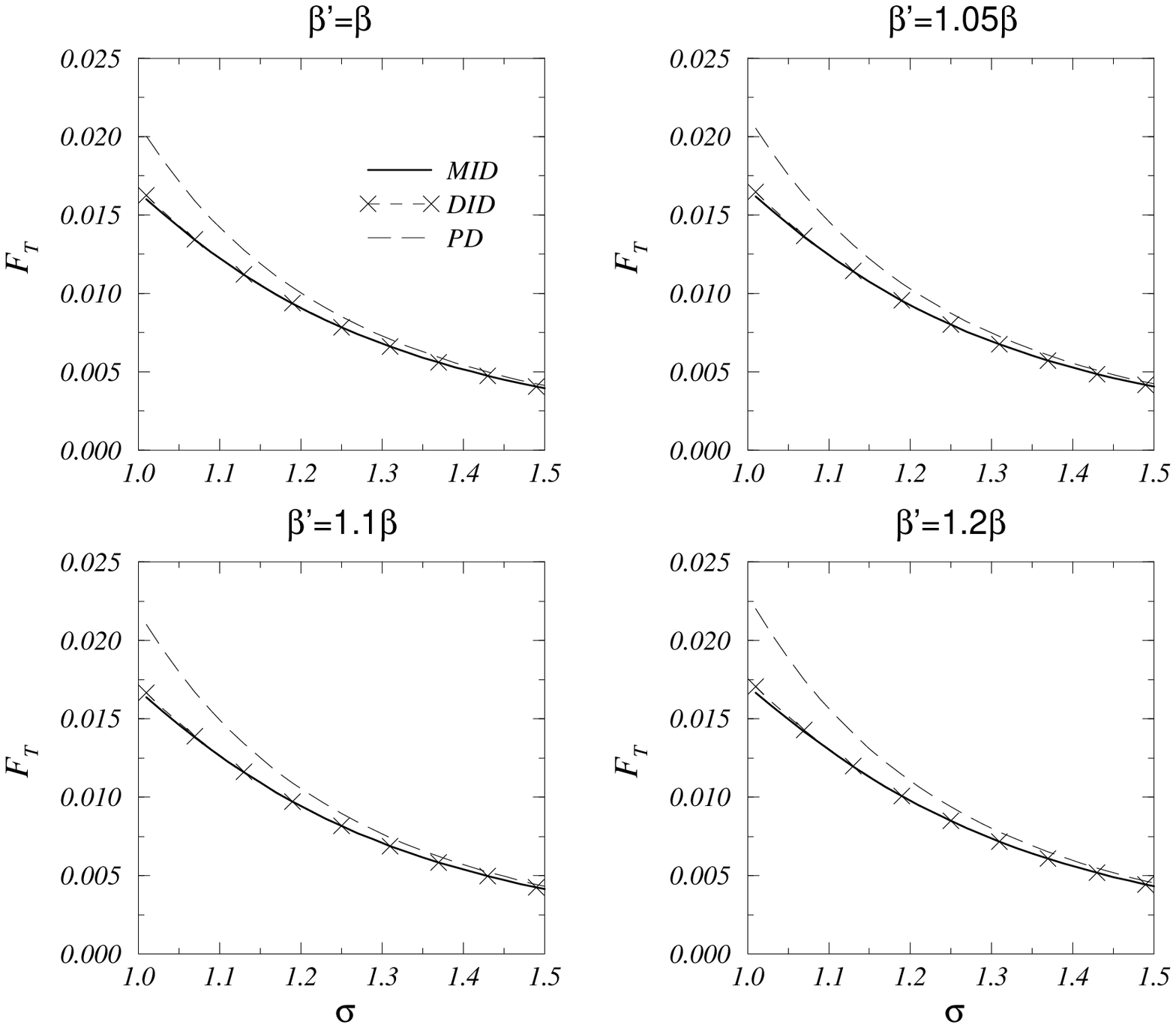,width=\linewidth}}

\newpage
\centerline{Fig.3: Force for Longitudinal Field, $\beta$=1/3}
\centerline{\epsfig{file=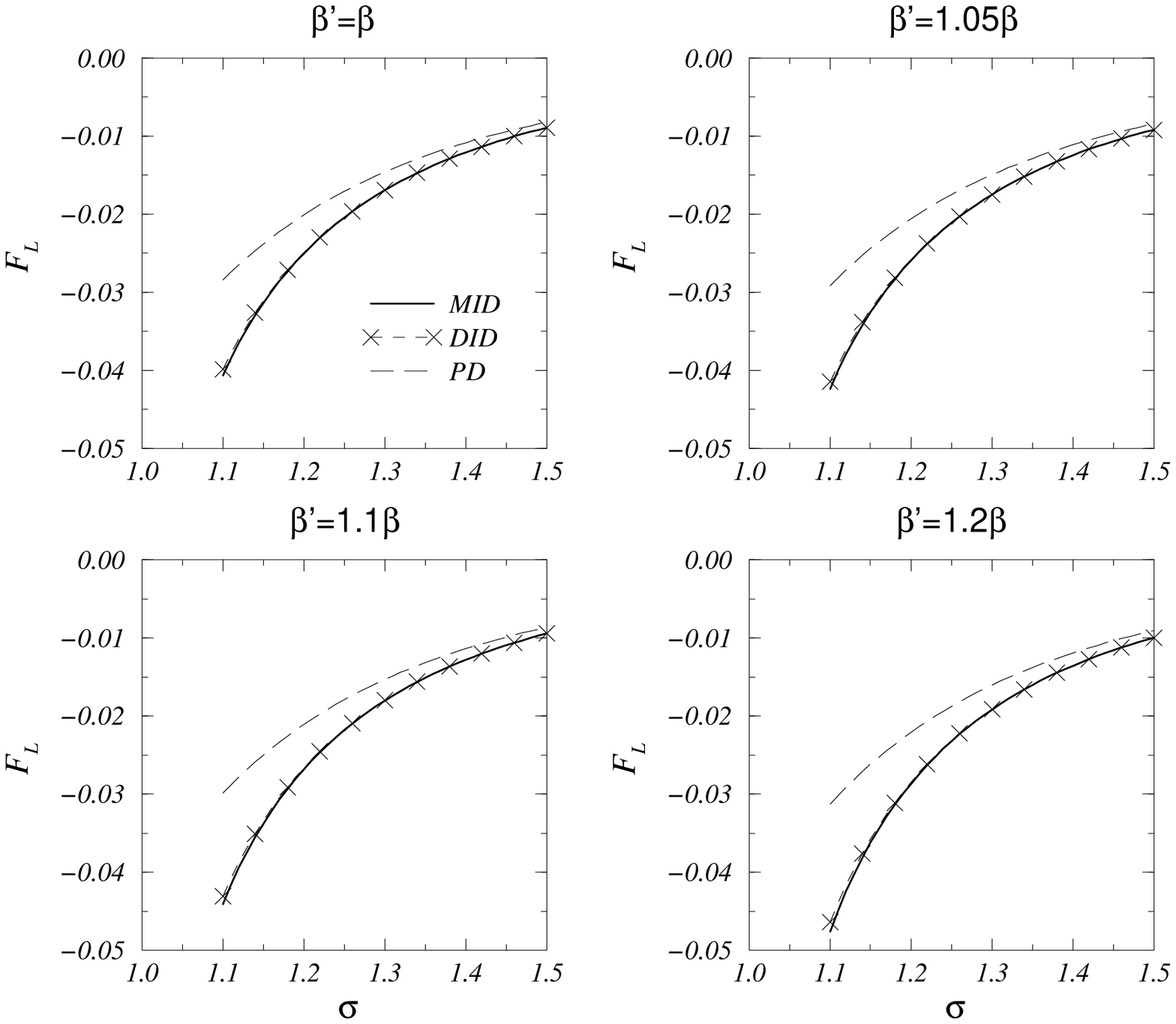,width=\linewidth}}

\newpage
\centerline{Fig.4: Force for Transverse Field, $\beta$=9/11}
\centerline{\epsfig{file=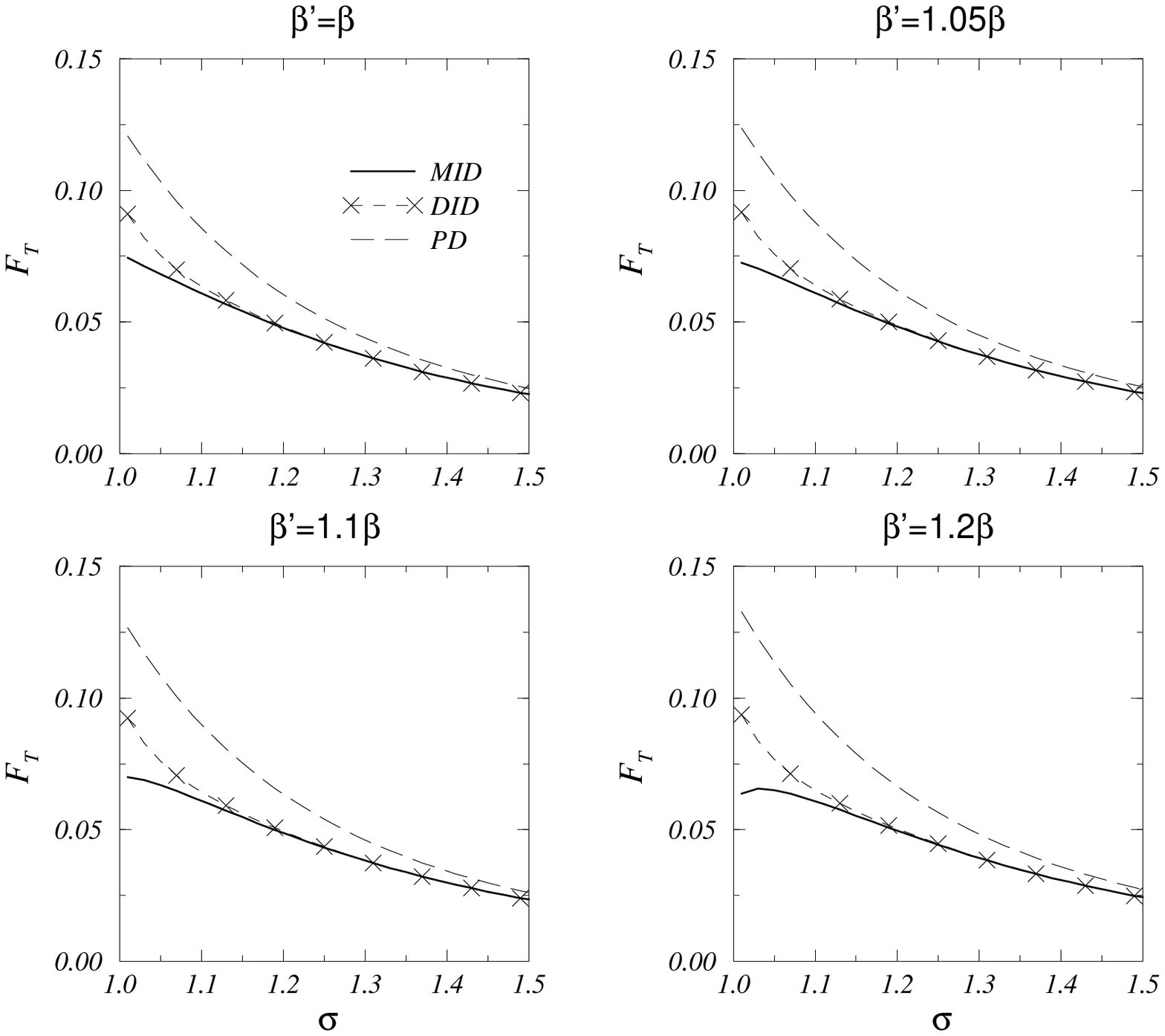,width=\linewidth}}

\newpage
\centerline{Fig.5: Force for Longitudinal Field, $\beta$=9/11}
\centerline{\epsfig{file=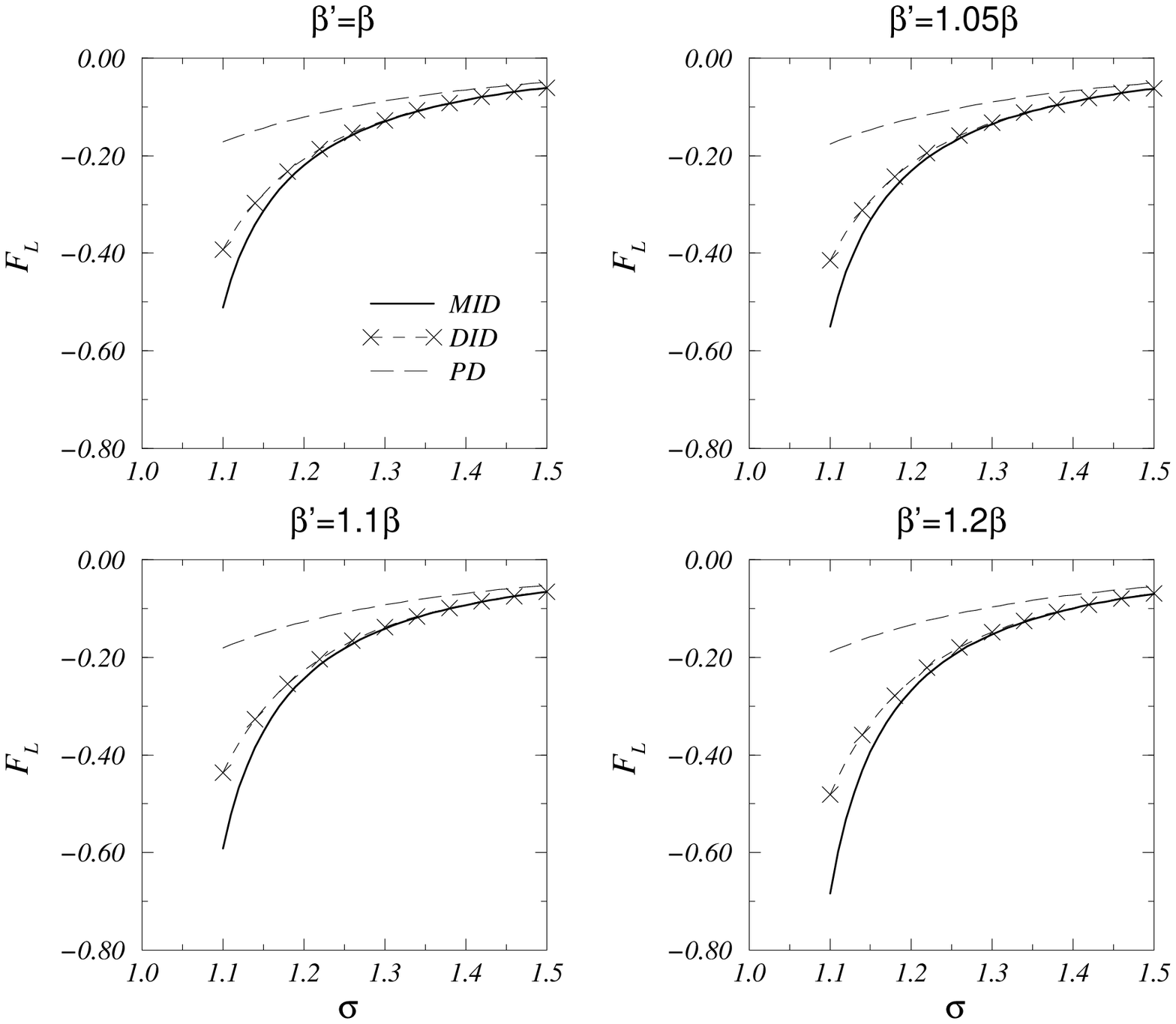,width=\linewidth}}

\end{document}